# Experimental exploration over a quantum control landscape through NMR

**Qiuyang Sun[1], István Pelczer[1], Gregory Riviello[1], Re-Bing Wu[2] and Herschel Rabitz[1]**

*1 Department of Chemistry, Princeton University, Princeton, New Jersey 08544, USA*

*2 Department of Automation, Tsinghua University and Center for Quantum Information Science and Technology, TNList, Beijing, 100084, China*

## Abstract

The growing successes in performing quantum control experiments motivated the development of control landscape analysis as a basis to explain these findings. When a quantum system is controlled by an electromagnetic field, the observable as a functional of the control field forms a landscape. Theoretical analyses have revealed many properties of control landscapes, especially regarding their slopes, curvatures and topologies. A full experimental assessment of the landscape predictions is important for future consideration of controlling quantum phenomena. NMR is exploited here as an ideal laboratory setting for quantitative testing of the landscape principles. The experiments are performed on a simple two-level proton system in a $H_2O$-$D_2O$ sample. We report a variety of NMR experiments roving over the control landscape based on estimation of the gradient and Hessian, including ascent or descent of the landscape, level set exploration, and an assessment of the theoretical predictions on the structure of the Hessian. The experimental results are fully consistent with the theoretical predictions. The procedures employed in this study provide the basis for future multi-spin control landscape exploration where additional features are predicted to exist.

## 1 Introduction

The control of quantum phenomena is garnering increasing interest for fundamental reasons as well as potential applications. Quantum system optimal control seeks to meet a posed physical objective, such as selective breaking of chemical bonds, creation of particular molecular vibrational excitations, manipulation of electron transfer in biological complexes, etc. (see [1] for a review). The control objective is generally addressed through the introduction of tailored electromagnetic fields, whose behavior is honed for the particular application. A practical means for the discovery of optimally designed fields is through the use of closed-loop learning control in the laboratory [2], where the objective signal measurement is fed back to a learning algorithm to evaluate the success of a candidate control field for refinement until the objective is reached as best as possible. The generally successful outcomes of optimal control experiments, as well as the extensive simulations on model systems, indicate that it is relatively easy to find good solutions while searching through the vast space of possible control fields. Seeking a fundamental explanation of this good fortune motivated the development of quantum control *landscape* analysis [3, 4, 5] which, upon satisfaction of particular assumptions, provides quantitative predictions on the nature of landscape features that can be tested in the laboratory. Many of the theoretical predictions have been successfully tested in large numbers of simulations [6, 7], but experimental affirmation is important for fundamental and practical reasons. For laser field manipulation of atomic, molecular or condensed phase systems, various laboratory complexities can make quantitative testing of landscape features challenging in these settings. Nuclear magnetic resonance (NMR) is best known as a powerful tool for chemical analysis, but for our purposes, the advanced nature of NMR instrumentation producing high signal-to-noise ratios (S/N) provides an ideal laboratory setting for testing the predictions from control landscape analysis. Control concepts have long been exploited experimentally in NMR [8, 9, 10, 11, 12, 13, 14, 15], and in this work we will utilize these capabilities for an initial study of control landscape principles in the most basic case of a two-level single-spin system. Although working with two or more coupled spins is needed to fully test the landscape analysis predictions [4, 5], the simple case of a single spin naturally must first be assessed.

In quantum control problems, like the case addressed here, the system evolution is governed by a time-dependent Hamiltonian, which is a function of a pulsed control field $C(t)$. The amplitude, phase, and/or frequency of the field can be modulated to meet the control objective, i.e., maximizing or minimizing an experimental measurable $J$ at the target time $T$. The objective may be expressed as $J{=}\mathrm{Tr}[\rho(T)O]$, where $O$ is a Hermitian observable operator and the density matrix $\rho(T)$ depends on the control $C(t)$ [4]. The functional dependence of $J[C(t)]$ upon the control field forms a control landscape. The time interval $T$ is chosen to be sufficiently long to permit unfettered control, while being short enough to consider the spin dynamics to form a closed system.

The topology of the landscape is important for determining the effectiveness of searches seeking an optimal control. A location on the landscape where the gradient satisfies $\delta J/\delta C(t) = 0$, $\forall t \in [0, T]$, specifies a *critical point.* An analysis of the landscape Hessian $\mathcal{H}(t, t') \equiv \delta^2 J/\delta C(t)\delta C(t')$ eigenvalues can reveal whether the critical point is a local extremum or saddle point. The number of positive and negative Hessian eigenvalues identify the intrinsic topological character at a critical point [5]. A critical point is locally maximal (or minimal) if the Hessian is negative (or positive) semidefinite, while a saddle point will have an indefinite Hessian spectrum.

The conclusions of control landscape analysis rest on the assumptions that (i) the system is controllable [16], (ii) the control to final state map, $C(t) \mapsto \rho(T)$, is surjective [17] and (iii) the controls are unconstrained [18]. Upon satisfaction of these assumptions, control landscape analysis predicts that the critical points only exist at particular values of the objective $J$ and that there are no local suboptimal maxima or minima (traps) over the landscape [4], thereby providing a basis to understand the observed relative ease of searching for optimal controls both in simulations [6, 7] and in the laboratory [1]. Although assumptions (i) and (ii) can be violated, the latter reported results suggest that they may be commonly satisfied. Assumption (iii) is always a concern in the laboratory where control resources are inevitably limited. The primary issue is whether the assumptions are satisfied to a practical degree in order to give good quality control performance.

The Hessian at a critical point has a specific maximum number of positive and negative eigenvalues dependent on the system's Hilbert space dimension, the initial density matrix $\rho(0)$ and the nature of the observable operator $O$ [5]. The latter Hessian non-zero eigenvalues are accompanied by an infinite dimensional null space. The present paper illustrates the most basic example of a control landscape for a proton spin-1/2 two-level system. Upon satisfaction of the three assumptions, this case is predicted to have no critical points except at the global maximum and mininum, and the Hessian at the landscape top (bottom) should possess at most two negative (positive) eigenvalues, respectively. This ideal control landscape picture from theoretical analysis might be altered by experimental imperfections such as relaxation through interaction with the environment and constraints on the control field [18], whose detailed impacts on the landscape structure remains an open challenge to assess. Extensive simulations have affirmed the predicted control landscape topology (see e.g., [6, 7]), and a recent laser experiment on pure state transitions in atomic rubidium was also consistent with expectations [19]. The present work demonstrates a set of experimental tools capable of roving over quantum control landscapes, including for future experiments in systems of interacting (coupled) spins and other circumstances.

The physical process underlying NMR can be viewed as the manipulation of magnetization vectors (proportional to the spin angular momentum) with pulsed magnetic fields in the radiofrequency (RF) regime as controls. NMR provides a desirable domain for studying fundamental properties of quantum control landscapes for several reasons: (i) Most nuclear spin systems have relatively simple and well-defined Hamiltonians. An isolated single spin-1/2 nucleus only possesses two energy levels upon interaction with the magnetic field, so the Hilbert space of a $n$-spin system is $2^n$-dimensional. Most molecules considered for laser control are too complex to model precisely, while spin systems under NMR control can often be modeled reliably with optimal control theory (OCT) as well as being readily amenable to the performance of optimal control experiments (OCEs) (see Fig.1). Thus, the option exists to either implement an OCT designed pulse shape in the laboratory to increase the efficiency of OCE, or utilize the laboratory OCE optimized pulse in a simulation to study the detailed controlled spin dynamics. (ii) Spin systems are well approximated as closed when the control interval $T$ is much shorter than both the longitudinal (system-environment) relaxation time ($T_1$) and the transverse (spin-spin) relaxation time ($T_2$) [8]. (iii) NMR machines are at an advanced stage of engineering capable of producing high fidelity shaped control pulses, and the observed spectra generally have high S/N of $10^3 \sim 10^4$, thus providing good accuracy for the measured observable. (iv) At conventional magnetic field strengths the Zeeman splitting is much smaller than $k_{\mathrm{B}}T$ at room temperature. Therefore, at thermal equilibrium a spin system is in a mixed state, which enables testing the general landscape topology predictions [4, 5]. The collective features (i)-(iv) above indicate that NMR provides a very attractive regime to test control landscape principles. Importantly, as these principles have a generic foundation, the implications of the findings extend beyond NMR to the control of other classes of systems using lasers or other sources.

Optimal control of nuclear spin systems has been treated theoretically and experimentally [9, 10, 11, 12, 13, 14, 15, 20]. Generally the experiments are executed with either of two approaches: (a) perform OCT to design an optimal control pulse and apply it in the laboratory [9, 10, 11]; (b) directly perform

OCE by iterative enhancement of a spectral signal to determine the control pulse [12, 13, 14, 15]. Both approaches have achieved success in the optimization of polarization transfer or dipolar decoupling in two-spin systems. However, these works have not addressed the analysis of the control landscape features, which is the focus of this paper. With regard to Fig.1, in this work OCT was performed in parallel with the OCE studies in order to provide a full understanding of the observations.

The remainder of the paper is organized as follows. Section 2 describes the methodology, including experimental determination of the gradient and Hessian and a series of algorithms as a basis to rove over the landscape. Section 3 gives the experimental setup for the two-level spin system in NMR (i.e., the proton in the HDO molecule in a mixture of 1% $H_2O$ and 99% $D_2O$), the objective $J$ and control variables. Experimental results are provided in Section 4 and a brief conclusion is given in Section 5.

## 2 Methods for experimental landscape roving

Experimental optimal control of quantum systems commonly use pulsed electromagnetic fields with adjustable amplitudes and phases over some spectral range as the input, and then the response signal from the physical system is recorded as the output. The various continuous control variables in the laboratory are inevitably discretized in some fashion, which we encapsulate here into a vector of $D$ control variables, $\vec{x} = (x_1, \cdots, x_D)^\mathsf{T}$. The focus of this work is on control landscapes, and the gradient and Hessian characterize the local features around a particular point $J(\vec{x}_0)$ on the landscape for control $\vec{x}_0$. The landscape gradient and Hessian evaluated at $\vec{x}_0$ respectively correspond to the vector $\nabla J$,

$$\nabla J(\vec{x}_0) = \left( \frac{\partial J}{\partial x_1}, \cdots, \frac{\partial J}{\partial x_D} \right)^\mathsf{T} \Bigg|_{\vec{x}_0}, \tag{1}$$

and the matrix $\mathcal{H}$ with elements

$$\mathcal{H}_{ij}(\vec{x}_0) = \mathcal{H}_{ji}(\vec{x}_0) = \frac{\partial^2 J}{\partial x_i \partial x_j} \Bigg|_{\vec{x}_0}, \qquad i, j = 1, \cdots, D. \tag{2}$$

We will construct "Rover" algorithms for systematically exploring the landscape utilizing the gradient and/or Hessian, whose experimental determination in the present two-level spin system will be described in Section 2.2.

## 2.1 The laboratory landscape Rover algorithms

In this work we introduce the concept of a control landscape Rover, which results from executing a suitable algorithm for taking an exploratory trajectory over the landscape in the laboratory. A trajectory can be characterized by a progress parameter $s \geq 0$, i.e., $\vec{x}(s)$, and the corresponding objective value $J[\vec{x}(s)]$. The trajectory roving over the landscape can be described by an ordinary differential equation

$$\frac{\mathrm{d}\vec{x}(s)}{\mathrm{d}s} = \vec{F}[\vec{x}(s)], \tag{3}$$

where the form of $\vec{F}$ is dictated by the particular landscape exploration goal. Four basic choices for $\vec{F}$ are given below.

The landscape roving operations in this work are constructed from four elementary operations, which can be understood as (i) continued movement along a fixed specified direction in the control space, (ii) performance of steepest ascent or descent of the landscape, (iii) horizontal movement on the landscape that preserves a non-critical value of $J$, and (iv) horizontal movement at a critical value of $J$. Arbitrary roving of the landscape may be performed by interleaving these elementary procedures as desired, with various possible goals including categorization of local landscape features beyond those of critical points.

- *Movement along a fixed specified direction.*

The movement in this case consists of "marching" in the same fixed direction $\vec{c}$ in the space of controls,

$$\frac{\mathrm{d}\vec{x}(s)}{\mathrm{d}s} = \vec{c}, \tag{4-i}$$

such that $\vec{x}(s) = \vec{x}(0) + s\vec{c}$. The choice of $\vec{c}$ is application specific, but a simple circumstance arises in the three cases below when the function $\vec{F}$ in Eq.(3) for vertical or horizontal movement (c.f., Eqs.(4-ii), (4-iii) or (4-iv)) is treated as a constant over a significant domain of the landscape.

- *Movement vertically.*

Here the movement is guided by the gradient for ascent or descent of the landscape,

$$\frac{\mathrm{d}\vec{x}(s)}{\mathrm{d}s} = \alpha \nabla J[\vec{x}(s)], \tag{4-ii}$$

where $\alpha$ is a scale factor. By solving Eq.(4-ii) for $\vec{x}(s)$, $J$ will increase ($\alpha > 0$) or decrease ($\alpha < 0$) monotonically until a critical point is reached.

- *Movement horizontally over a non-critical level set.*

Although extremizing $J$ is a common physical goal, exploring all landscape features is important for assessing the basic theory. Thus, a third case is horizontal movement on the landscape, or level set exploration [21, 22] corresponding to continous roving without changing the value of $J$. A quantum control *level set* is defined by a family of control solutions which all achieve the same objective value $J$, and the distinct members over the level set may show large variation of secondary characteristics. Simulations have explored non-critical level sets (i.e., away from where $\nabla J = \mathbf{0}$) [21] as well as the landscape top/bottom [22]. A trajectory on a non-critical level set is characterized by the local change in the control $\mathrm{d}\vec{x}(s)$ being orthogonal to the gradient $\nabla J[\vec{x}(s)]$. In the $D$-dimensional control space, there are $(D-1)$ linearly independent directions orthogonal to the gradient, so the level set trajectory from a specified starting point is not unique. The following differential equation satisfies these conditions:

$$\frac{\mathrm{d}\vec{x}(s)}{\mathrm{d}s} = \vec{g} - \nabla J(\nabla J^{\mathsf{T}} \cdot \vec{g})/(\nabla J^{\mathsf{T}} \cdot \nabla J), \tag{4-iii}$$

where $\vec{g}$ is an arbitrary vector of length $D$ whose choice will guide the level set trajectory.

- *Movement horizontally over a critical level set.*

Equation (4-iii) does not apply at landscape critical points, such as the top or bottom, where the gradient is zero. In this case the level set trajectory lying in the null space of the Hessian will keep the value of $J$ invariant at the critical point,

$$\frac{\mathrm{d}\vec{x}(s)}{\mathrm{d}s} = \sum_i \vec{v}_i(s) \left[\vec{v}_i^{\mathsf{T}}(s) \cdot \vec{h}\right], \tag{4-iv}$$

where $\{\vec{v}_i\}$ are the eigenvectors of the Hessian with zero eigenvalue, and $\vec{h}$ is an arbitrary vector of length $D$ whose choice guides the critical level set exploration trajectory. In the extreme case that the Hessian had no zero eigenvalues, the critical point would exist in isolation on the landscape forbidding movement as the r.h.s. of Eq.(4-iv) would be zero.

Eqs.(4(i-iv)) are special cases of Eq.(3) which will be used during the experiments reported in Section 4. Since the landscape gradient and Hessian at an arbitrary point can be measured in the laboratory, the differential equations can be solved in real time while performing the ongoing experiments. In this work the forward Euler method was found to be sufficient,

$$\vec{x}(k+1) = \vec{x}(k) + \beta\vec{F}[\vec{x}(k)], \qquad k = 0, 1, \cdots \tag{5}$$

where $\vec{x}(k)$ is the control in the $k$-th iteration (i.e., the $k$-th step of $s$) and $\beta$ is the step size. In other applications, especially when the S/N is not high, statistical averaging of the data at each step and higher order integration methods may be needed.

## 2.2 Gradient and Hessian determination

Experimental determination of the gradient and Hessian of objective $J$ in this work is based on making small increments about a current control $\vec{x}_0$ and then measuring the resultant changes in the associated $J$ values. For the landscape gradient we found that a simple central difference method was stable,

$$\frac{\partial J}{\partial x_i} \approx \frac{J(\cdots, x_i + d_i, \cdots) - J(\cdots, x_i - d_i, \cdots)}{2d_i}, \qquad i = 1, \cdots, D, \tag{6}$$

where $d_i$ is a small increment of the variable $x_i$ which should be reasonably chosen based on the nature of $x_i$ and $J$ in a particular experiment. Similarly, the second-order partial derivatives in the Hessian also can be standardly expressed with finite differences. However, this method has a high S/N requirement for measuring $J$ because of noise sensitivity, and we found that estimation of the Hessian was problematic by direct application of finite differences. Statistical strategies can be employed to reliably extract quality gradients and Hessians from experimental data [19, 23]. In this work we utilize least squares (LS) to determine the Hessian from the data $J(\vec{x}_0 + \Delta\vec{x})$ with a set of perturbations $\Delta\vec{x}$. For this purpose the landscape can be approximated about $\vec{x}_0$ by a second-order Taylor series

$$J(\vec{x}_0 + \Delta\vec{x}) \approx J(\vec{x}_0) + \nabla J(\vec{x}_0)^{\mathsf{T}} \cdot \Delta\vec{x} + \frac{1}{2}\Delta\vec{x}^{\mathsf{T}} \cdot \mathcal{H}(\vec{x}_0) \cdot \Delta\vec{x}. \tag{7}$$

Determining $\mathcal{H}(\vec{x}_0)$ also requires extracting the gradient $\nabla J(\vec{x}_0)$, and the linear system in Eq.(7) has $D(D+3)/2$ unknowns: $D$ for the gradient and $D(D+1)/2$ for the Hessian. With sufficient random samples of $\Delta\vec{x}$ 's about $\vec{x}_0$ (i.e., typically ~500 samples were used in this work) the overdetermined linear system can be solved with LS to obtain the Hessian. Although the gradient is also determined in solving Eq.(7), when only the gradient was called for in landscape ascent or descent and for non-critical level set exploration, the more efficient central difference method in Eq.(6) was used.

## 3 Experimental setup

The dynamics of a two-level system, i.e., a nuclear spin-1/2 with gyromagnetic ratio $\gamma$, is quantum-mechanically formulated as follows. A closed quantum system in a mixed state is described by its density matrix $\rho(t)$, whose time evolution is governed by the Hamiltonian $H(t)$ according to the Liouville-von Neumann equation,

$$\frac{d}{dt}\rho(t) = -\frac{i}{\hbar}[H(t), \rho(t)]. \tag{8}$$

In the case of a two-level spin system, a static and homogeneous magnetic field $B_0$ is implemented; its orientation defines the $z$-axis for the nuclear magnetization both in the laboratory and in the rotating frame. A shaped RF control pulse is applied orthogonal to $B_0$ (i.e., in the $x$-$y$ plane) with a carrier frequency of $\omega_{\mathrm{RF}} > 0$. The pulse has adjustable amplitude $A(t)$ and phase $\phi(t)$, which are usually slowly varying functions in the time domain compared with the carrier frequency. In a rotating frame about the $z$-axis at frequency $\omega_{\mathrm{RF}}$, whose orientations of $+x$ and $+y$ are defined with the phase angles of $0°$ and $90°$, respectively, we conveniently define two new functions in terms of $A(t)$ and $\phi(t)$ to specify a control pulse, i.e., $B_x(t) := A(t)\cos\phi(t)$ and $B_y(t) := A(t)\sin\phi(t)$. This formulation enables us to express the full Hamiltonian in the rotating frame as [8]

$$H(t) = -(\gamma B_0 - \omega_{\mathrm{RF}})I_z - \gamma[B_x(t)I_x + B_y(t)I_y], \tag{9}$$

where $I_x$, $I_y$ and $I_z$ are the spin angular momentum operators, related to the Pauli matrices as $I_j = \hbar\sigma_j/2$, $j = x, y, z$. Although for special choices of $B_x(t)$ and $B_y(t)$ Eq.(8) may be solved analytically, here we seek to establish the capability of freely roving over the landscape and testing the landscape predictions which generally requires full freedom in the field forms.

The NMR experiments presented in this paper were implemented on a Bruker Avance-III 500 MHz spectrometer, equipped with a TCI ($^1$H/$^{13}$C/$^{15}$N/$^2$H) cryoprobe and highly digitized and linear RF signal generator (SGU) (Bruker-Biospin, Billerica, MA). We used a sealed standard 1%$H_2O$/99%$D_2O$ sample which contains a small amount of $GdCl_3$ to accelerate the $T_2$ relaxation processes. The measured values for $T_1$ and $T_2^*$ (which includes local inhomogeneity effects as well) are 187 ms and 70 ms, respectively. In this mixture the rapid exchange and the overwhelming excess of $D_2O$ assures that the dominant species available for $^1$H-detection is HDO. The H/D coupling is effectively removed by the rapid exchange as well (i.e., exchange decoupling [24]). All these features permit treating the sample as a two-level system consisting of a single spin with a singlet resonance. All the experiments were performed at 295K after careful manual tuning and shimming of the magnet. The carrier frequency $\omega_{\rm RF}$ was set exactly on-resonance, thus the $I_z$ term of the Hamiltonian in Eq.(9) can be dropped. In this work each shaped pulse with a fixed final time of $T = 500\mu s$ is broken into four equal time intervals of constant field value, so the control is given as a vector of length $D = 8$, i.e., $\vec{x} = (B_x^1, \cdots, B_x^4, B_y^1, \cdots, B_y^4)^\intercal$, where $B_x^i$ ($B_y^i$) is the value of the corresponding control within the $i$-th time interval. The set of eight variables is an approximation to the true control as a freely varying continuous function of time, and in the present experiments this choice of discrete variables proved to be adequate for satisfying resource assumption (iii) of control landscape analysis. With the above experimental setup each measurement takes ∼3 sec of laboratory time, including the pulse duration, data acquisition time of ∼0.5 s, and ∼2.5 s waiting for the system to relax back to its equilibrium state.

The control objective considered here is to steer the initial state $\rho(0) \propto I_z$ under thermal equilibrium to $-\langle I_x \rangle$ with a suitably shaped pulse. The objective $J = -\langle I_x \rangle$ is characterized by the integrated area of the corresponding singlet peak in the NMR spectrum. The zeroth-order phase correction [25] is calibrated manually so that the peak area represents the magnetization in the desired orientation $-x$, and then fixed throughout the experiment; the first-order phase correction [25] is always set to zero. If the maximal peak integral is $J_{\rm max}$, then the full domain of $J$ will be $[-J_{\rm max}, J_{\rm max}]$ since the peak can be either positive or negative depending on the control pulse. To determine the level of noise we took 100 repeated measurements of several typical control pulses producing various $J$ values on the landscape; in each case the error approximately obeyed a Gaussian distribution with a standard deviation of $(10^{-4} \sim 10^{-3})J_{\rm max}$.

# 4 Results and discussion

The laboratory experiments aimed to demonstrate the landscape roving capability while also assessing the particular predictions of landscape analysis. Starting from an arbitrarily chosen initial control field, we first performed optimization by ascent and descent of the landscape to reach the top and bottom in Section 4.1, resulting in a full landscape trajectory. The Hessian spectrum was evaluated at the top and bottom of the landscape in Section 4.2, along with performance of an extensive excursion through the space of controls while roving on the top of the landscape. Section 4.3 presents two non-critical level set exploration experiments optimizing a secondary characteristic to find the associated best control upon traversing a landscape level set.

## 4.1 Ascending and descending the landscape

Landscape roving starts with gradient ascent or descent, i.e., maximization or minimization of the objective $J$ by moving the control along the local gradient measured in each iteration, until either halting at a suboptimal critical point trap or reaching the full landscape top/bottom extrema. Arbitrary units are used for the objective and control fields below, and their respective scales will be consistent throughout Section 4 (e.g., a constant 90° pulse with a length of $T = 500\mu s$ corresponds to a field strength of ∼37). Fig.2(a) shows a typical optimization curve of $J$ containing ∼100 iterations. The randomly chosen initial control corresponds to the 0th iteration, which was then optimized for both ascent and descent on the landscape to give the full curve. The Euclidean norm of the experimentally measured landscape gradient at each iteration is also displayed. As the search approached the landscape maximum or minimum, $J$ converged and continued to randomly vary in a narrow range with $J_{\min} \simeq -J_{\max}$ within the level of noise, while the gradient norm converged toward zero. The gradient norm curve is more noisy than that of $J$, especially in the near-optimal regions, due to noise amplification. The control landscape analysis predicts that upon satisfaction of the three underlying assumptions, there should be no suboptimal traps or saddle points on such a two-level system landscape [4], which is confirmed by the experiments.

The control field at the initial iteration as well as at the landscape maximum and minimum in the trajectory are shown in Fig.2(b). When a magnetic field interacts with a spin-1/2 proton it will rotate the magnetization vector clockwise about the magnetic field direction. During maximization and minimization of the control objective $\langle I_z \rangle \to -\langle I_x \rangle$, the $B_x$ component of the control gradually diminishes while $B_y$ becomes dominant, providing the desired rotation of the initial magnetization along the $+z$ axis toward the desired orientation along $x$. Note that there are infinitely many optimal solutions for this control problem, the simplest among which is just a DC field in the $+y$ orientation of the rotating frame with a flip angle of 90° to give the maximal value of $J$. However, in the trajectory we present here even at the landscape optimum the $B_x$ component is not zero. Once the initial control $\vec{x}(0)$ is specified, the gradient ascent trajectory $\vec{x}(s)$, $s \geq 0$ is deterministic and only one optimal control is found. Thus, level set exploration is employed below to access other portions of the landscape.

### 4.2 Hessian observation over the landscape and exploration of the landscape top

The Hessian matrix gives an assessment of the curvature at any location on the landscape. In particular, the landscape analysis for a two-level system predicts a specific rank of at most 2 for the Hessian at the top and bottom critical points [5], which can be tested in the laboratory. To further assess the Hessian character we evaluated it using least squares (see Section 2.2) at five selected locations, $J/J_{\max}$=1.00, 0.71, 0.31, 0.03, and -1.00, along the gradient ascent/descent search trajectory in Fig.2. Approximately 500 small random control sampling variations were performed about each of the five points to determine the associated Hessian matrices with their eigenvalues shown in Fig.3. In the middle region of the landscape, the Hessian has positive and negative eigenvalues with comparable magnitudes. As $J$ approaches the landscape top, the Hessian spectrum moves toward being negative semidefinite. At the maximum point ($J/J_{\max}$=1.00), two Hessian eigenvalues are significantly negative while the remaining six are zero within experimental error, in agreement with the theoretical prediction. Similarly, the Hessian spectrum at the landscape minimum $J/J_{\max} = -1.00$ has two positive eigenvalues with the remaining six being zero within the level of noise, again agreeing with theory [5]. Simulations were performed considering the laboratory noise level, confirming the statistical quality of the results.

The Hessian eigenvectors associated with non-zero eigenvalues at a maximum (minimum) point of the landscape describe the independent paths for driving off the landscape top (bottom) region, and the magnitudes of the eigenvalues characterize the sensitivity of $J$ to variation along these paths in the control space. Similarly, the eigenvectors with zero eigenvalue specify directions for remaining on the top (bottom) of the landscape. The two eigenvectors $\vec{v}_1$ and $\vec{v}_2$ determined from the measured Hessian at the maximum point $J/J_{\max}$=1.00 are presented in Fig.4, corresponding to the eigenvalues $\lambda_1 = -316$ and $\lambda_2 = -150$, respectively. The eigenvectors show that $J$ can be decreased by perturbing the optimal control in two independent coordinated ways: (i) adding an approximate constant field to the $B_y$ component expressed by $\vec{v}_1$, (ii) adding a specially shaped field to $B_x$ with a particular weaker variation of $B_y$, expressed by $\vec{v}_2$.

As an assessment of the landscape structure about an extremum point [19], straight roving was performed with Eq.(4-i) along each of the Hessian eigenvectors in the vicinity of the initial critical point. Using the landscape maximum at $\vec{x}_0$ shown in Fig.2(b) as a starting point, we moved along each of the eight eigenvectors with the successive choice of $\vec{c} = \vec{v}_i$, $i = 1, \cdots, 8$ in Eq.(4-i). The process was characterized by the relative distance of a control $\vec{x}$ from the starting point $\vec{x}_0$ in the control space, i.e., $\|\vec{x} - \vec{x}_0\|/\|\vec{x}_0\|$, as shown in Fig.5. A parabolic drop of $J$ was observed when moving along the directions specified by either of the two eigenvetors $\vec{v}_1$ and $\vec{v}_2$. In contrast, $J$ remained almost constant while moving along the other six eigenvectors over appreciable roving distances of at least $\pm 100\%$, indicating that the Hessian null space has broad extent. Importantly, through the choice of the vector $\vec{h}$ in Eq.(4-iv) motion along the top of the landscape also can be expressed by a wandering trajectory specified as an arbitrary linear combination of the Hessian null space eigenvectors.

Building on the results in Fig.5 and the comment above, we implemented a trajectory to extensively "drive" over the top of the landscape by employing Eq.(4-iv), i.e., remeasuring the Hessian about the critical point in each iteration and moving in the null space of the local Hessian. Again we started from the maximum point found in Section 4.1 and chose a simple constant free vector $\vec{h} = (1, 1, 1, 1, 1, 1, 1, 1)^\intercal$ to obtain the control trajectory on the top of the landscape given in Fig.6. The Hessian in each iteration was estimated by the LS with 100 random samples, whose spectrum was always found to be dominated by two significantly negative eigenvalues, and the other six Hessian eigenvalues being essentially zero. The objective value stayed very close to the landscape maximum throughout the exploration, and the vector of control variables evolved from the starting point by a relative distance of more than 240% at the final iteration. These results again suggest that the top of the landscape is quite extensive, although no attempt was made to fully explore its scope.

### 4.3 Non-critical objective value level set exploration

Section 4.2 examined the critical level set at the top of the landscape, while here we consider non-critical exploration at $J \neq J_{\max}$ or $J_{\min}$. As described by Eq.(4-iii), traversal of a non-critical level set requires movement along a path that is locally orthogonal to the gradient $\nabla J$ on the landscape. In practice, this is performed by choosing any free vector $\vec{g}$ in Eq.(4-iii) with the projection operation on the r.h.s. removing the component of $\vec{g}$ along the gradient. A randomly varying $\vec{g}$ may lead to a random walk in the level set, but here we will take advantage of the freedom in choosing $\vec{g}$ by a specification that optimizes a secondary characteristic goal $K[\vec{x}(s)]$ .

As a first example, we consider minimizing the energy $K_E$ of the control pulse:

$$K_E[\vec{x}(s)] \propto \int_0^T \left[B_x^2(t) + B_y^2(t)\right] \mathrm{d}t \propto \sum_{i=1}^{4} \left[(B_x^i)^2 + (B_y^i)^2\right] = \|\vec{x}(s)\|^2. \tag{10}$$

Minimizing the pulse energy can be achieved by choosing the vector $\vec{g}(s)$ proportional to the derivative of the energy function, i.e.,

$$\vec{g}(s) = -\frac{\vec{x}(s)}{\|\vec{x}(s)\|} \tag{11}$$

The weight $\|\vec{x}(s)\|$ in the denominator scales $\vec{g}$ to the local magnitude of the control. The experimental results in Fig.7(a) start with an arbitrarily chosen initial control producing $J_0 = 0.585 J_{\max}$ and then a level set trajectoy is taken while minimizing $K_E(\vec{x})$. The pulse energy decreased by more than 50% over $\sim$20 iterations, while the $J$ value varied in a narrow range of $\sim 0.003 J_{\max}$. The error bars in Fig.7(a) for $J/J_{\max}$ were determined from the standard deviations of five repeated measurements at each iteration. The fluctuation of the level set value was comparable with the random noise in measuring $J$, thereby showing good stability for the level set exploration. The evolution of the control field over the level set is given in Fig.7(b). The results show that the energy-optimal control pulse on the level set is converging towards a simple DC field in $y$ direction.

As a second example of non-critical level set exploration, we desire to move as far as possible from the starting point in the control space. Therefore, we consider the secondary characteristic $K_d$ of the Euclidean distance squared from the initial control $\vec{x}(0)$:

$$K_d[\vec{x}(s)] = \|\vec{x}(s) - \vec{x}(0)\|^2, \tag{12}$$

and maximization of $K_d$ can be achieved by defining the free vector proportional to the derivative of $K_d$, i.e.,

$$\vec{g}(s) = \vec{x}(s) - \vec{x}(0). \tag{13}$$

The free vector is randomly chosen at the initial iteration to start the exploration, and after that it is determined by Eq.(13). In order to move a long distance on a landscape level set efficiently in the laboratory where measuring the gradient is time-consuming, we modestly sacrifice the precision of the level set. Thus, we adopt the procedure of continuing to exploit the current gradient direction as far as possible during the exploration in order to reduce the frequency of gradient measurement. We move along that constant direction combining Eqs.(4-i) and (4-iii) until the drop from the original level set value $J_0$ exceeds a specified tolerance, and then we measure the gradient and correct the control by gradient ascent or descent to bring it back to the original level set value $J_0$ followed by further level set exploration, etc. The step size of the gradient correction is determined as follows. According to the gradient ascent algorithm in Eqs.(4-ii), we obtain

$$\frac{\mathrm{d}J}{\mathrm{d}s} = \frac{\partial J}{\partial \vec{x}(s)} \cdot \frac{\mathrm{d}\vec{x}(s)}{\mathrm{d}s} = \alpha \left\| \frac{\partial J}{\partial \vec{x}(s)} \right\|^2 \equiv \alpha \left\| \nabla J[\vec{x}(s)] \right\|^2 . \tag{14}$$

A gradient correction of $J$ is performed with Eq.(5) (where $\beta = \alpha$ now) as follows at the $k$-th iteration $\vec{x}(k)$ in order to return $J$ back to $J_0$:

$$\vec{x}(k+1) = \vec{x}(k) - \frac{J - J_0}{\|\nabla J\|^2} \nabla J \bigg|_{\vec{x}(k)} \tag{15}$$

with the coefficient $-(J - J_0)/\|\nabla J\|^2$ assuring ascent or descent as needed and conservative movement by the normalization in the denominator. Therefore, the full landscape level set algorithm is a combination of horizontal and vertical restoring elementary movements as follows:

1) Specify a non-critical initial control $\vec{x}(0)$, measure the objective value $J_0$ and the gradient at $\vec{x}(0)$, and randomly choose a free vector $\vec{g}$ for Eq.(4-iii);

2) Determine the roving direction by removing from $\vec{g}$ its projection on the gradient through the operation on the r.h.s. of Eq.(4-iii);

3) Move the control $\vec{x}$ along the level set roving direction by a constant distance (thus the coefficient $\beta$ in Eq.(5) is non-constant in this example) combining Eqs.(4-i) and (4-iii), and measure the new $J$;

4) If $|J - J_0| \leq \epsilon$, where $\epsilon$ is the tolerance for deviation from the original level set, go to step 3);

If $|J - J_0| > \epsilon$, measure the gradient and correct the control according to Eq.(15), then measure the new $J$ at the corrected control. When $|J - J_0| < \epsilon$ is obtained, reset the free vector $\vec{g}$ by Eq.(13) and then go to step 2);

5) Iterate until $\vec{x}(s)$ has evolved a specified distance $\sqrt{K_d}$ in the control space from $\vec{x}(0)$.

Figure 8(a) shows a distance-maximization level set roving experiment using the algorithm above, where the tolerance for being acceptably close to $J_0$ in order to define a level set is chosen to be $|J - J_0| \leq 0.014 J_{\max}$. Over the set of ∼30 iterations the gradient correction was performed only three times, and after each correction $J$ returned to its initial value $J_0$ (dashed line in the figure) with good precision. By the end of the roving trajectory the control vector had changed by a relative distance of >250%, while taking only less than 20 min of laboratory time. The controls at several selected iterations, i.e., the initial and final ones as well as before and after each gradient correction, are shown in Fig.8(b). We see that the corrections only caused relatively small variations of the control while returning to the original $J_0$ value, and the overall procedure did not break the continuity of the level set exploration to the specified tolerance.

# 5 Conclusion

This work reported the first experimental study of an NMR control landscape including extensive roving. The high S/N in NMR experiments enabled the determination of the gradient and Hessian at any point over the landscape with good precision and efficiency. With knowledge of the local gradient and/or Hessian, we implemented various landscape roving algorithms serving specific purposes, including optimizing the objective $J$, finding controls on a non-critical level set while optimizing the value of a secondary characteristic and exploration of the top of the landscape. The theoretical predictions on the landscape topology were verified with good accuracy. The methodology developed in this paper can be further exploited to assess the control landscape analysis in more complex coupled multi-spin systems, which are expected to have saddle points at particular values of $J$ with specified Hessian signatures. The findings of such landscape topology assessment studies are of interest in spin systems under control, but the implications extend beyond for the control of other quantum mechnical physical phenomena with electromagnetic fields.

## Acknowledgement

The authors acknowledge support from the NSF Grant No. CHE-1058644, DOE Grant No. DE-FG02-02ER15344, and USC/ARO/MURI Grant No. 160018. R.-B. Wu acknowledges support from the NSFC Grant No. 61374091 and 61134008.

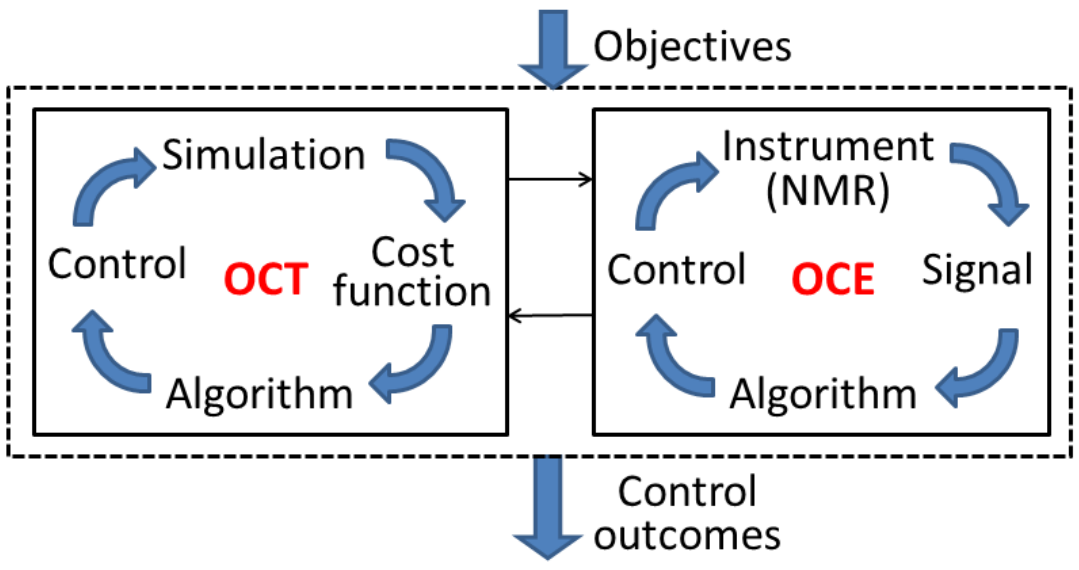

Figure 1: (Color online) The relationship between the dual loops of optimal control theory (OCT) and optimal control experiments (OCE) in NMR systems. Both control loops share the common features of implementing a field, calculating or measuring the cost function, assessing how well the objective is achieved, and varying the control iteratively guided by an algorithm. The high quality of spin system Hamiltonians permits balanced operation with OCT and OCE, as called for by the control objectives. In the present work on assessing control landscape analysis predictions, the experiments were exclusively performed with OCE and the findings subsequently affirmed in OCT simulations. In more complex scenarios with multiple spins the performance of OCT may provide reasonable control estimates to reduce the level of experimental effort.

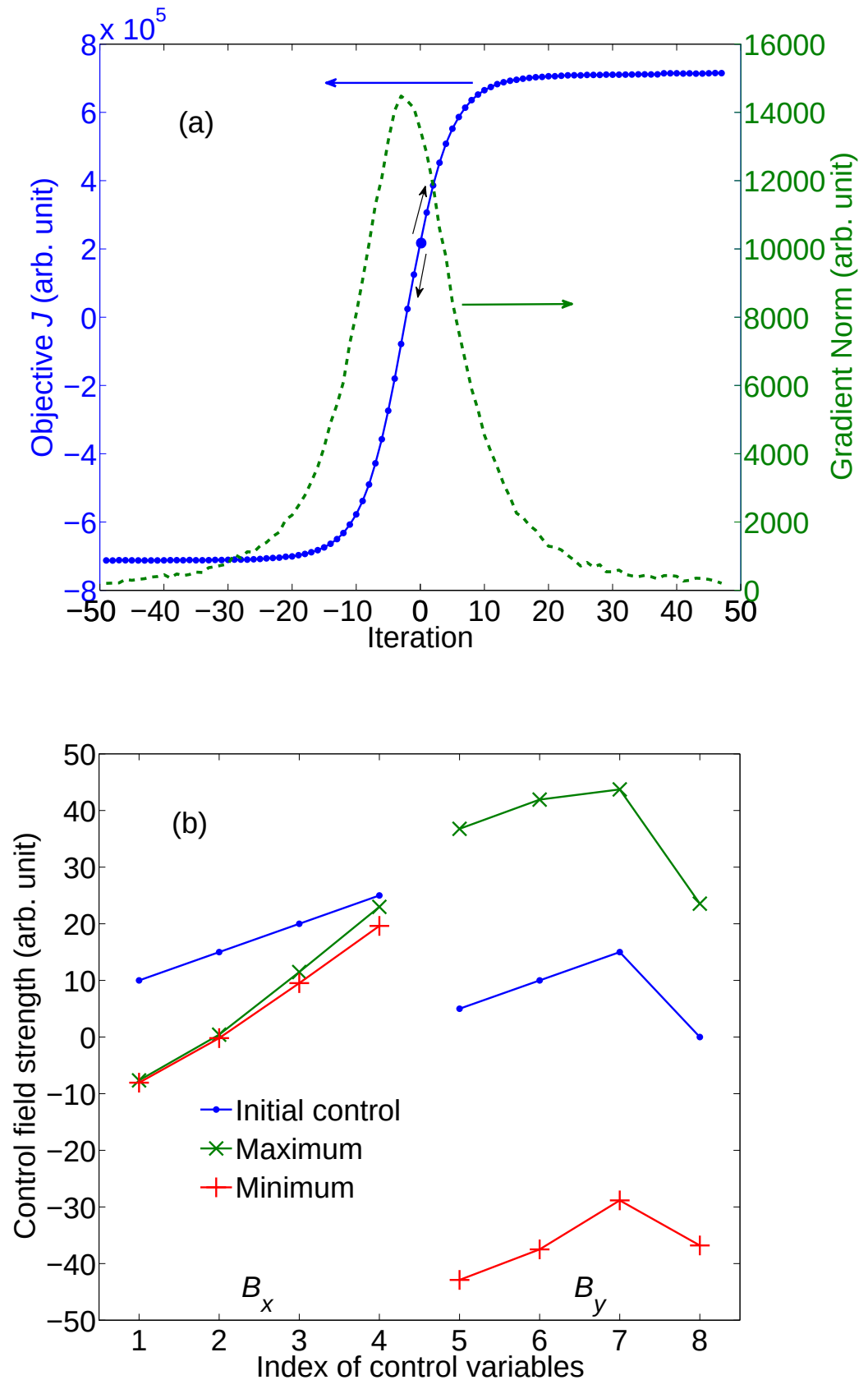

Figure 2: (Color online) Landscape ascent/descent along the directions speficified by the gradient of $J$ (the peak integral of the $^1$H NMR spectrum in HDO). (a) The evolution of the objective $J$ and the norm of its gradient $(\nabla J^\intercal \cdot \nabla J)^{1/2}$ during optimization. The initial condition corresponds to the 0-th iteration shown as a bold dot for $J$; the iterations proceed in the positive (negative) direction to ascend (descend) the landscape. (b) The control pulse shapes at the landscape maximum and minimum points in the above search trajectory, compared with the randomly picked initial control.

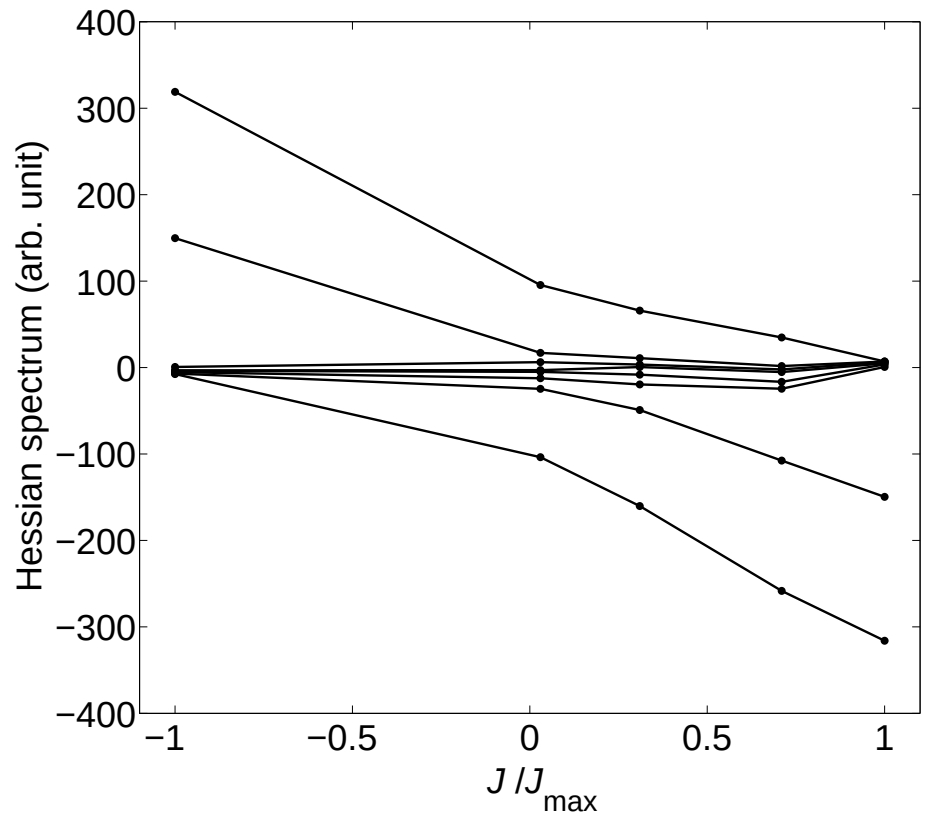

Figure 3: The Hessian spectrum evaluated at five selected values of $J/J_{\max}$ along the gradient ascent (descent) search in the trajectory of Fig.2. The spectrum swinges from being negative semidefinite at $J/J_{\max} = 1.00$ to being positive semidefinite at $J/J_{\max} = -1.00$. The existence of two ($\pm$) non-zero eigenvalues at the landscape bottom (top) is consistent with theoretical predictions.

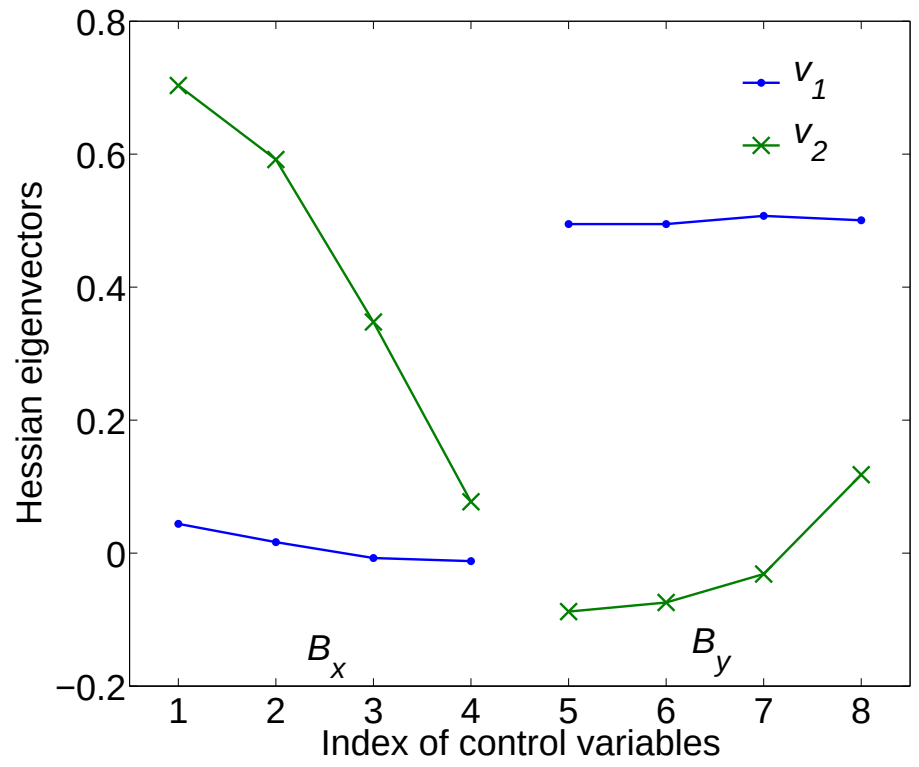

Figure 4: (Color online) The two Hessian eigenvectors $\vec{v}_1$ and $\vec{v}_2$ respectively corresponding to the two dominant eigenvalues $\lambda_1$ and $\lambda_2$ at the landscape top ($J/J_{\max}$=1.00).

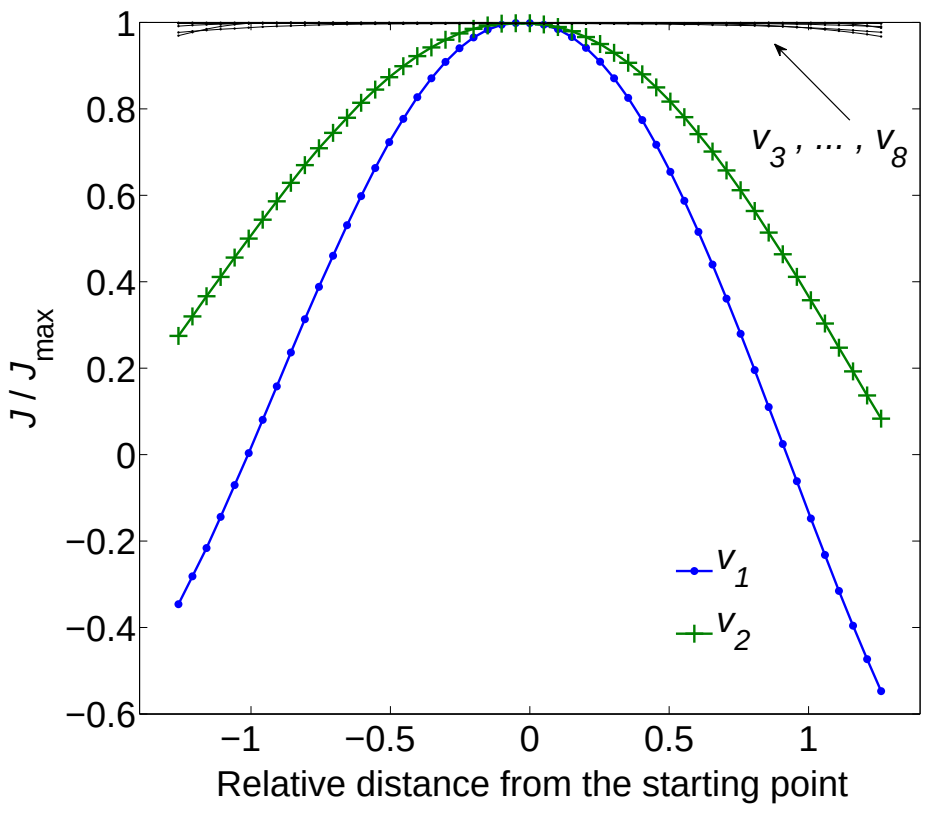

Figure 5: (Color online) Driving on and off the landscape top by proceeding along directions specified by each of the eight Hessian eigenvectors at the starting point. Continued marching along the directions specified by either of the two eigenvectors with negative eigenvalues takes $J$ down the landscape along nearly parabolic paths. In contrast, marching along any of the six eigenvectors with approximately zero eigenvalue leaves $J \simeq J_{\max}$ while the control accordingly changes form. In all cases, movement along any of the eigenvectors by relative distances even greater than $\pm 100\%$ still preserved the physical meaning of the eigenvectors reflected in their eigenvalues (i.e., being either negative or zero).

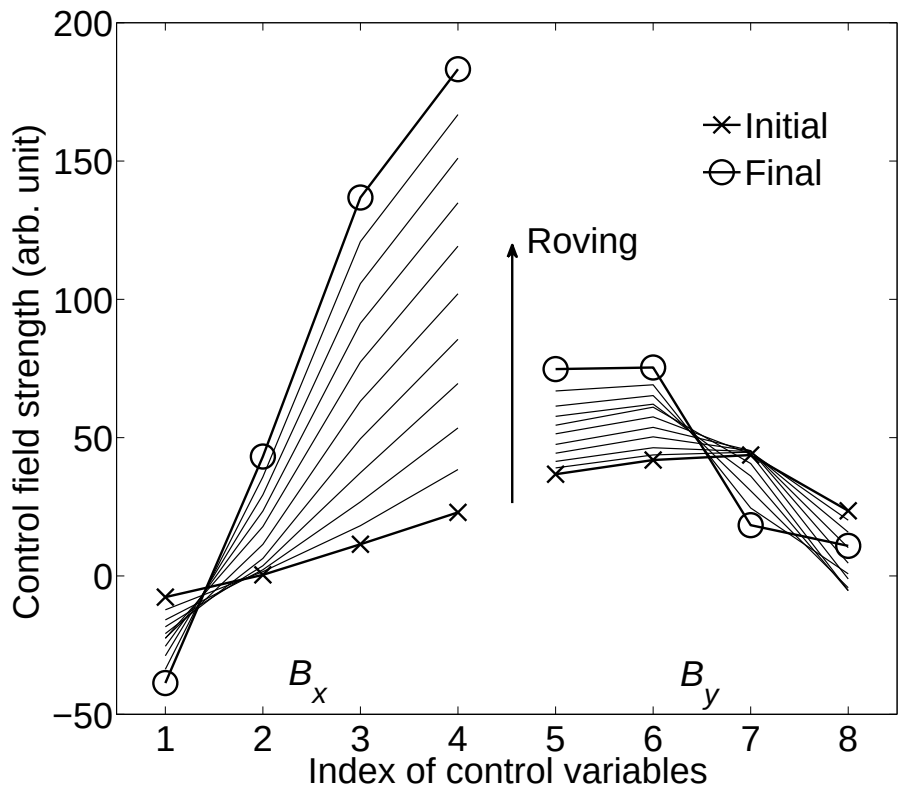

Figure 6: Shapes of the evolving control field in a 10-iteration exploration of the landscape top by roving iteratively in the Hessian null space utilizing controls given by solving Eq.(4-iv).

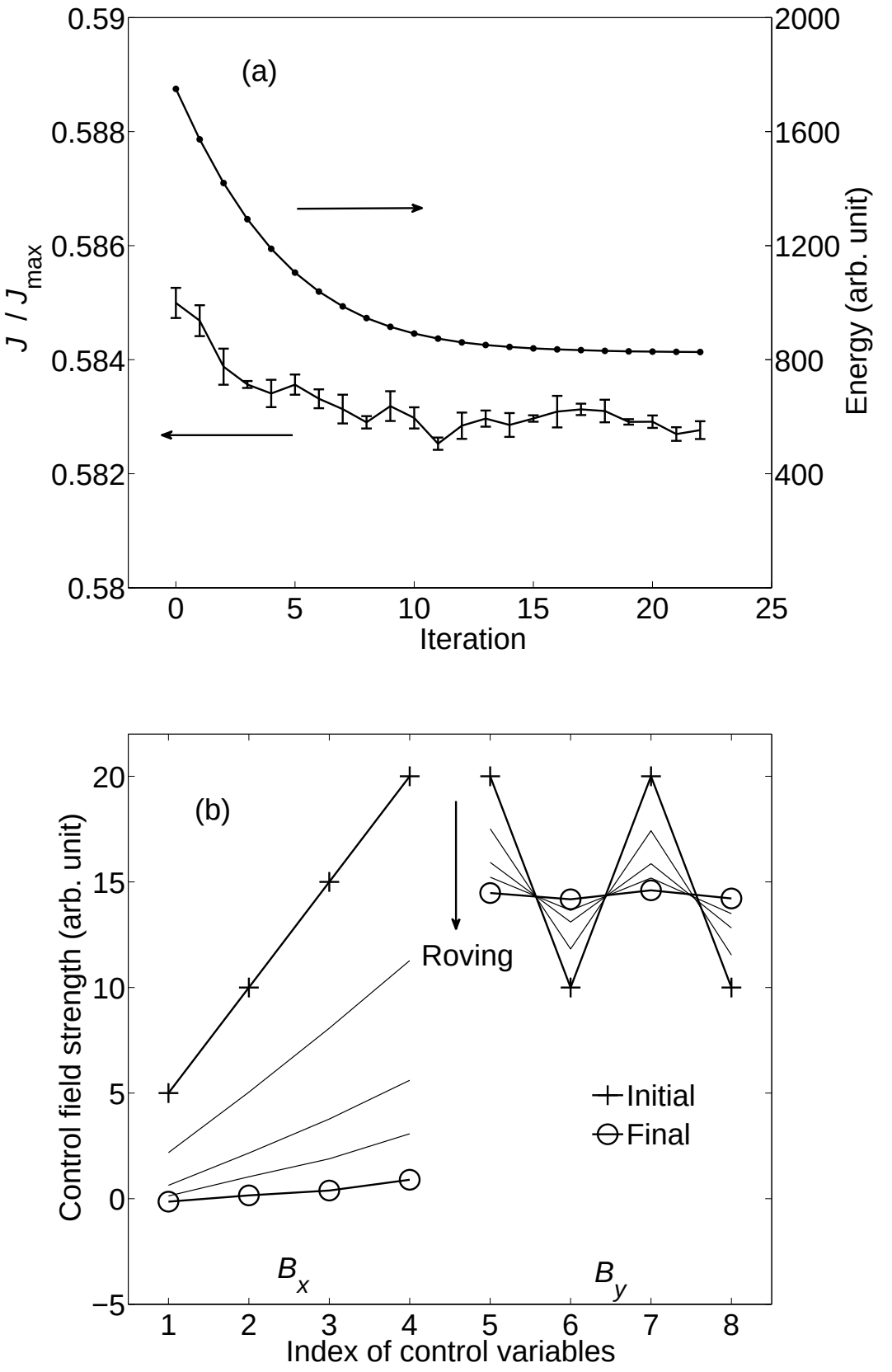

Figure 7: An experiment for energy-minimization on a non-critical point level set exploration: (a) The pulse energy drops significantly by over 50% at the end of the excursion while $J$ remains nearly constant showing a variation of less than $0.3\% J_{\max}$. (b) The controls at iterations 0 (initial), 5, 10, 15, and 22 (final) in (a), showing evolution of the control during the level set exploration.

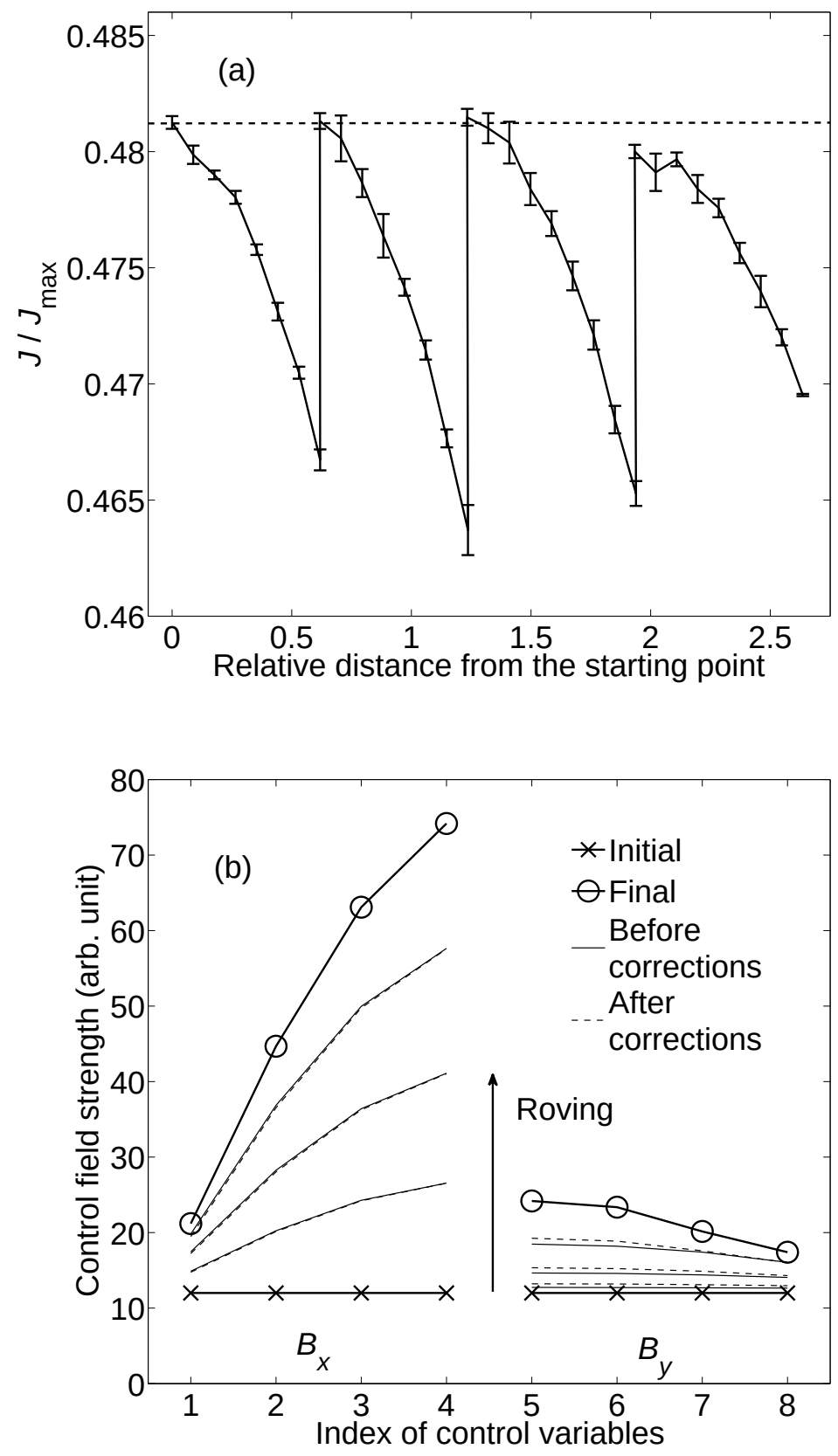

Figure 8: A landscape non-critical level set roving experiment aiming to move as far away as possible from an initial control: (a) Evolution of the objective $J$; the three sudden jumps correspond to gradient ascent corrections when the deviation of $J$ from the initial value $J_0 = 0.482 J_{\max}$ exceeded the tolerance of more than $0.014 J_{\max}$. The horizontal dashed line shows the position of the targeted level set. (b) The control fields at the initial and final iterations, as well as before (after) each gradient correction shown by the three groups of solid (dashed) lines. The experiment was terminated when the roving field evolved by a relative distance of $\sim$250%.